\documentclass[5p]{elsarticle}

\usepackage{amsmath}
\usepackage{booktabs}

\usepackage{graphicx}
\usepackage{subfigure}
\usepackage{float}
\usepackage{hyperref}
\usepackage[all]{hypcap}

\journal{Mechanics Research Communications}

\begin{document}

\begin{frontmatter}

\title{Massively-Parallel Lagrangian Particle Code and Applications}

\author[SBU]{Shaohua Yuan}
\author[SBU]{Mario Zepeda Aguilar}
\author[SBU]{Nizar Naitlho}
\author[SBU,BNL]{Roman Samulyak\corref{Roman Samulyak}}
\cortext[Roman Samulyak]{Corresponding author}
\ead{roman.samulyak@stonybrook.edu}

\address[SBU]{Department of Applied Mathematics and Statistics, Stony Brook University, Stony Brook, New York 11794, USA.}
\address[BNL]{Computational Science Initiative, Brookhaven National Laboratory, Upton, New York 11973, USA.}

\begin{abstract}
Massively-parallel, distributed-memory algorithms  for the Lagrangian particle hydrodynamic method [R. Samulyak, X. Wang, H.-C. Chen, Lagrangian particle method for compressible fluid dynamics, J. Comput. Phys.,  362 (2018), 1-19] have been developed, verified, and implemented. The key component of parallel algorithms is a particle management module that includes a parallel construction of octree databases, dynamic adaptation and refinement of octrees, and particle migration between parallel subdomains. The particle management module is based on the p4est (parallel forest of k-trees) library.  The massively-parallel Lagrangian particle code has been applied to a variety of fundamental science and applied problems. A summary of Lagrangian particle code applications to the injection of impurities into thermonuclear fusion devices and to the simulation of supersonic hydrogen jets in support of laser-plasma wakefield acceleration projects has also been presented.     

\end{abstract}


\end{frontmatter}

\tableofcontents

\section{Introduction}

Particle-based Lagrangian methods \cite{Monaghan2005,Frontiere2017,Avesani2014} possess a number of unique features that make them the ideal choice for numerical simulations of certain classes of problems. By discretizing the medium (compressible fluid, gas, plasma etc.) with particles, Lagrangian methods achieve a continuum, high-level of adaptivity: particles move with the flow increasing their density in compressed regions and automatically provide refinement / coarsening of numerical resolution where it is needed. Such a continuous adjustment of resolution eliminates the presence of interfaces between fine and coarse mesh patches, present in adaptive mesh refinement, that may create a source of numerical artifacts. 
Lagrangian particle methods are capable of working with extremely non-uniform distributions of matter (as typical in astrophysics and some high-energy-density applications) as they do not need to resolve "empty"  space between sparse regions of matter. Particle methods can resolve material boundaries or interfaces of high geometric complexity and they exactly conserve certain properties of matter in numerical simulations. Like all numerical methods, they should be used in their best applicability range. For example, they would under-perform compared to certain spectral methods if a very high resolution is necessary in a rectangular domain with periodic boundary conditions.

Smooth particle hydrodynamics (SPH) was the 1st Lagrangian particle method suitable for 3D computational fluid dynamics. The traditional SPH method \cite{Monaghan1992} was often limited by a major drawback: its very poor accuracy of discrete differential operators. It is widely accepted \cite{Monaghan2005,Diltz1999,Hopkins2014} that the traditional SPH discretization has zeroth-order convergence for widely used kernels. Numerous methods have been developed to improve the SPH discretization of differential operators. They include the Reproducing Kernel Particle Methods (RKPM) \cite{Liu1995} and the Normalized Smoothing Function (NSF) methods \cite{Oger2007}, the Conservative Reproducing Kernel Smoothed Particle Hydrodynamics (CRKSPH) \cite{Frontiere2017}, and moving-least-squares WENO-SPH schemes \cite{Avesani2014}.

While most of the previous works present modifications and improvements of the SPH while still preserving the main ideas of SPH and its smooth kernel, the Lagrangian particle (LP) method \cite{SamulyakWang2018} significantly deviates from them. It implements an unsplit Lax–Wendroff type scheme with a limiter. In the location of points selected by a state estimator, the Lax–Wendroff scheme is blended with a first-order upwind method to control numerical dispersion.
In the LP method, approximations of spatial derivatives are obtained by employing a local polynomial fit known also as the generalized finite difference (GFD) method \cite{BenitoUrena2001}, which is always accurate to a prescribed order. The LP method is second-order accurate in space and time and it is extendable to higher order accuracy. In addition to compressible hydrodynamics, the LP code implements equations of low magnetic Reynolds number magnetohydrodynamics \cite{SamulyakYuan2021} and granular flow equations \cite{Zepeda2022}.

The 1st version of the Lagrangian particle code \cite{SamulyakWang2018} was written for multi-core processors that significantly limited its applicability to large scale scientific simulations. In this paper, we describe a parallel Lagrangian particle algorithm and the corresponding object-oriented code for massively-parallel, distributed memory supercomputers. The massively - parallel Lagrangian particle code is being used for a variety of fundamental science and applied problems, in particular for the simulation of impurities in thermonuclear fusion devices. In the last Section, we briefly summarize the thermonuclear fusion applications and describe in more detail hydrodynamic simulations for the nozzle design optimization for supersonic hydrogen jets. Such simulations support experimental work at Brookhaven National Laboratory on laser-plasma interaction and laser-driven plasma wakefield acceleration. 


\section{\label{sec:models}Lagrangian Particle models and algorithms}

\subsection{\label{sec:equations}Main governing equations} 

The core Lagrangian particle (LP) method \cite{SamulyakWang2018} solves the system of compressible 3D Euler equations in Lagrangian coordinates:
\begin{eqnarray}\label{euler_eq}
	&&\frac{d \rho}{d t}  = - \rho \nabla \mathbf{u}, \label{eq:mass}\\
	&&\rho \frac{d \mathbf{u}}{dt}= -\nabla P, \label{eq:momentum}\\
	&&\rho \frac{de}{dt} = -P\nabla \cdot \mathbf{u},\label{eq:energy}\\
	&& P = P(\rho,e), \label{eq:eos}
\end{eqnarray}
where $d/dt$ is the Lagrangian time derivative, $\mathbf{u}$, $\rho$ and $e$ are the velocity, density and specific internal energy, respectively, and $P$ is the pressure. In addition, various extensions of the system (\ref{eq:mass} - \ref{eq:eos}), including magnetohydrodynamic equations in the low Magnetic Reynolds number approximation \cite{SamulyakYuan2021} and granular flow equations \cite{Zepeda2022}, have been implemented. The LP method also implements various models for the equation of state (\ref{eq:eos}) describing gases, plasmas, and liquids. 

The LP method can be summarized as follows. The medium is discretized by Lagrangian particles, i.e. each Lagrangian particle represents a fluid parcel characterized by density, pressure, internal energy and other quantities. The number density of Lagrangain particles is proportional to the mass density of the medium at each point in space. Differential operators are discretized  by employing a local polynomial fit known also as the generalized finite difference (GFD) method \cite{BenitoUrena2001}. The main component of the GFD method, finding neighborhoods for every Lagrangian particle in a highly non-uniform system, is implemented via octree data structure construction and search \cite{Meagher1982}. A second-order accurate in space and time discretization of the system is obtained based on the Lax-Wendroff method. The second order method, while used most of the time in numerical simulations, is prone to numerical dispersion in the location of discontinuities and high gradients of physics states which leads to unphysical oscillations of states. To eliminate numerical dispersion, the 2nd order method is blended with a 1st order accurate, diffusive upwind method via a limiter in locations detected by a state estimator. After updating states of the Lagrangian particles, they are moved to the next location in space by the velocity field. These main components of the Lagrangian particle method are described below.

The main idea of the 1st order upwind method is easy to introduce for a 1D system of Euler equations which can be written in the following quasi-linear form using $U=[V$ $u$ $P]^{T}$ as the state vector:
\begin{equation} \label{euler3}
	U_{t} + A(U)U_{x} = 0,
\end{equation}
\begin{equation} \label{euler4}
	U = \left( \begin{array} {c} V \\ u \\ P \end{array} \right), \quad
	A(U) = V\left( \begin{array} {ccc} 
		0 & -1 & 0 \\ 
		0 & 0 & 1  \\ 
		0 & K & 0 \end{array} \right),
\end{equation}
where 
\begin{equation} \label{Kdef}
	K = \left(P+\frac{\partial e}{\partial V}\right)\left/\frac{\partial e}{\partial P}\right. ,
\end{equation}
and $V = 1/\rho$ is the specific volume.
Performing diagonalization of $A$ as $A = R\Lambda R^{-1}$, 
equations (\ref{euler3}) and (\ref{euler4}) become 
\begin{equation} 
	U_{t} + R\Lambda R^{-1}U_{x} = 0 \nonumber,
\end{equation}
\begin{equation} \label{euler5}
	R^{-1}U_{t} + \Lambda R^{-1}U_{x} = 0,
\end{equation}
where
\begin{equation} 
	R^{-1} = \left( \begin{array} {ccc} 
		1 & 0 & \frac{1}{K} \\ 
		0 & -\frac{1}{2 \sqrt{K}} & -\frac{1}{2K} \\ 
		0 & \frac{1}{2 \sqrt{K}} & -\frac{1}{2K} \end{array} \right), 
	\,\,
	R = \left( \begin{array} {ccc} 
		1 & 1 & 1 \\ 
		0 & -\sqrt{K} & \sqrt{K}  \\ 
		0 & -K & -K \end{array} \right) \nonumber,
\end{equation}
\begin{equation} \label{euler6}
	\Lambda = V\left( \begin{array} {ccc} 
		0 &          &   \\ 
		& \sqrt{K} &   \\ 
		&          &  -\sqrt{K} \end{array} \right)	
\end{equation}
This gives rise to three advection equations for components of $R^{-1}U$ which are solved by placing numerical stencils in the upwind direction. In terms of the original state quantities, the upwind equations are 
\begin{eqnarray}
	V_t &=& \frac{V}{2} \left( u_{xr}+u_{xl} \right) -
	\frac{V}{2 \sqrt{K}} \left( P_{xr}-P_{xl} \right), \label{eq_upwind_V}\\ 
	u_t &=& \frac{V \sqrt{K}}{2} \left( u_{xr}-u_{xl} \right) -
	\frac{V}{2} \left( P_{xr}+P_{xl} \right), \label{eq_upwind_u}\\
	P_t &=& -\frac{VK}{2} \left( u_{xr}+u_{xl} \right) + 
	\frac{V \sqrt{K}}{2} \left( P_{xr}-P_{xl} \right). \label{eq_upwind_P}
\end{eqnarray}
The subscripts $l$ and $r$ to the spatial derivatives
in equations (\ref{eq_upwind_V} - \ref{eq_upwind_P}) indicate that the corresponding terms, in the discrete form, will be computed using one-sided left (l) or right (r) derivatives. Extension of these equations to the 3D case can be found in \cite{SamulyakWang2018}.

The first-order upwind scheme is stable but diffusive. We now present a second-order, unspilt, Lax-Wendroff-type method based on symmetric in space stencils for Lagrangian particles. As before, we consider a 1D Lagrangian formulation of Euler equation for polytropic gas. Differentiating (\ref{euler3}) - (\ref{Kdef}) with respect to $t$ 
and exchanging the order of the material and spatial derivatives, we obtain
\begin{eqnarray}
	&&V_{tt}+VV_xP_x+V^2P_{xx}=0 \nonumber\\
	&&u_{tt}-\gamma Vu_xP_x-\gamma VPu_{xx}=0 \\
	&&P_{tt}-(\gamma^2+\gamma) P u_x^2-\gamma V_xPP_x-\gamma VPP_{xx}=0 \nonumber
\end{eqnarray}
Notice that the exchange of the order of the material and spatial derivatives introduces an additional term 
\begin{equation}
	\frac{Df_x}{Dt}=\left(\frac{Df}{Dt}\right)_x-u_xf_x.
\end{equation}
Using the Taylor series expansion, we obtain the semi-discrete equations
\[
\frac{V^{n+1}-V^n}{\Delta t}=Vu_x+\frac{1}{2}\Delta t (-VV_xP_x-V^2P_{xx})+O(\Delta t^2) \\
\]
\[
\frac{u^{n+1}-u^n}{\Delta t}=-VP_x+\frac{1}{2}\Delta t (\gamma VPu_{xx}-Vu_xP_x)+O(\Delta t^2) \\
\]
\[
\frac{P^{n+1}-P^n}{\Delta t}=-\gamma Pu_x+\frac{1}{2}\Delta t (\gamma^2 Pu_x^2+\gamma V_xPP_x+\gamma VPP_{xx}+
\]
\[
\gamma Pu_x^2) + O(\Delta t^2)
\]
The derivation in two- and three-dimensional spaces is tedious but straightforward. The corresponding equations are given in \cite{SamulyakWang2018}

Numerical approximations of spatial differential operators at locations of Lagrangian particles are obtained using the Generalized Finite Differences \cite{BenitoUrena2001}. Suppose we need to calculate derivatives of a function $U$ at the location of particle 0 in a 2D domain.  In the vicinity of particle $0$, the function value in the location of particle $i$ can be expressed by
the Taylor series as
\begin{equation}
	U_i = U_0+h_i\left.\frac{\partial U}{\partial x}\right|_0+k_i\left.\frac{\partial U}{\partial y}\right|_0+
\end{equation}
\[
\frac12\left(h^2_i\left.\frac{\partial^2 U}{\partial x^2}\right|_0 + k^2_i\left.\frac{\partial^2 U}{\partial y^2}\right|_0 + 2h_ik_i\left.\frac{\partial^2 U}{\partial x \partial y}\right|_0\right) + \ldots ,
\]
where $U_i$ and $U_0$ are the corresponding function values in the location of particles
$i$ and $0$,  $h_i = x_i - x_0$, $k_i = y_i - y_0$, 
and the derivatives are calculated in the location of particle $0$.
Let's truncate the Taylor series to a second-order polynomial.  
In order to approximate values of the derivatives,
we perform a local polynomial fitting using states on $m>=5$ particles in the vicinity of particle $0$.  The following linear system $Ax=b$
\begin{equation} \label{linear_system_axb}
	\begin{bmatrix} h_1 & k_1 & \frac12 h^2_1 & \frac12 k^2_1 & h_1k_1\\ 
		h_2 & k_2 & \frac12 h^2_2 & \frac12 k^2_2 & h_2k_2 \\  
		\vdots & \vdots & \vdots & \vdots & \vdots \\ 
		h_m & k_m & \frac12 h^2_m & \frac12 k^2_m & h_mk_m
	\end{bmatrix}  
	\left[ \begin{array}{c} 
		U_x \\ U_y  \\ U_{xx} \\ U_{yy} \\ U_{xy} 
	\end{array} \right] 
	=
	\left[ \begin{array}{c} 
		U_1 - U_0 \\ U_2 - U_0 \\ \vdots \\ U_m - U_0 
	\end{array} \right],
\end{equation}
is usually overdetermined.
An optimal solution to (\ref{linear_system_axb}) is a solution $x$ that minimizes the $L_2$ norm of
the residual, i.e.,
\begin{equation}
	min\|Ax-b\|_2,
\end{equation}
and the QR decomposition with column pivoting is employed to obtain $x$. 

For the second-order method, particle neighborhoods are intended to be symmetric while including particles from all 8 octants. For the upwind method, the left and right derivatives are approximated using the left and right semi-spheres in the corresponding spatial dimension. The 2nd-order method is blended with the upwind method as
\begin{equation}
	s_i^{n+1} = s_i^n + \Delta t \left( UW_i - \phi(r_i)\left( UW_i - LW_i\right) \right), 
\end{equation}
where $s_i^n$ is a hydrodynamic state at point $i$ at time $t^n$, $UW_i$ and $LW_i$ schematically denote the upwind and Lax-Wendroff method at point $i$, respectively, and $\phi(r_i)$ is a limiter function based on values of a state estimator. The state  estimator  takes into consideration the quotient of the left $s_{xl}$ and right $s_{xr}$ derivatives and is defined as
\[ r_i = \begin{cases} 
	\infty & s_{xl} \cdot s_{xr} \leq 0\\
	\frac{1}{2}\left( \frac{s_{xr}}{s_{xl}} + \frac{s_{xl}}{s_{xr}} \right)   & elsewhere 
\end{cases}
\]
The limiter controls the total variation of numerical solutions and reduces the numerical order of accuracy at extremum points and high solution gradients.
The specific form of the limiter function is often problem dependent. In some simulations, we use the following empirically designed limiter function
\[ \phi(r) = \begin{cases} 
	1.25 e^{(1-\sqrt{r})} & r < \frac{3}{2}\\
	1.25 e^{-\frac{9}{2}(\sqrt{r}-1)^2} & r > \frac{3}{2}  
\end{cases}
\]
that effectively eliminates the dispersion while optimizing the convergence rate.

After the update of states of each Lagrangian particle, particles are advanced in space by a mixture of the forward and backward Euler schemes:
\begin{eqnarray} \label{timeIntegrate}
	\frac{x^{n+1}-x^n}{\Delta t} &=& \frac{1}{2}\left(
	u^n+u^{n+1}\right)
\end{eqnarray}

Other details of the method can be found in \cite{SamulyakWang2018}.

\subsection{\label{sec:algorithms}Massively parallel algorithms} 

In this Section, we describe main algorithms included in the time step loop and their implementation for massively parallel supercomputers.
The fluid particles are generated during the initialization. The LP algorithm for each time step can be presented as the following steps for a 3D problem. 

1. \textbf{Boundary.} Generate boundary particles for given boundary conditions. For example, mirror particles are used for solid boundary conditions. Inflow and outflow boundaries can also be set up at this stage by properly generating or deleting particles.

2. \textbf{Octree.} Construct an octree for the present distribution of fluid and boundary particles. Adaptively refine or coarsen the octree until the particle number in an octant does not exceed a prescribed number.

3. \textbf{Neighbor searching.} Find  neighbors for each particles in six directions (top, down, front, back, left, right). If the free boundary condition is applied and neighbors are not complete in certain directions, ghost particles will be generated to complete the neighborhood.

4. \textbf{Stencil.} Given the neighbors of each particles, generate stencils for both the upwind and Lax-Wendroff methods. For the upwind method, neighbors are organized into six lists in order to compute spatial derivatives in six directions. Stencils are close to spherical in shape for the Lax-Wendroff method. A simple sorting of particles in the distance ascending order may lead to the selection of most particles from one direction while missing information in other directions. Thus, the second principle of ordering neighbor lists is that the top members of each list should contain neighbors from all directions.

5. \textbf{Solver.} Use GFD to compute spatial derivatives. The optimal number of neighbors for the GFD solver is obtained from numerical experiments. 8 and 54 neighboring particles are used in upwind and Lax-Wendroff schemes to compute spatial derivatives. Then we compute time derivatives based on the numerical discretization.

6. \textbf{Particle mover.} Update states of particles and move them to the next location in space.

The major octree algorithms such as building, refining and searching that significantly affect the code performance are parallelized using the p4est library. p4est \cite{BursteddeWilcoxGhattas11} enables a dynamic management of a collection of adaptive octrees on distributed memory supercomputers. It has the functionality of building, refining, coarsening, 2:1 balancing and partitioning on computational domains composed of multiple connected two-dimensional quadtrees or three-dimensional octrees, referred to as a forest of octrees. Here we briefly introduce three features of p4est which greatly contribute to the massively parallel LP code. 

\begin{figure}[h!]
	\centering
	\includegraphics[width=0.9\linewidth]{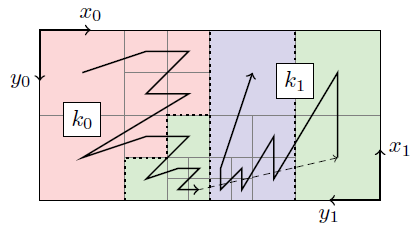}
	\caption{A 2D example with two octrees($k_0$ and $k_1$). A left-to-right traversal of leaves through two octrees creates a space-filling z-curve(black line). In this example the forest is partitioned among three processes(represented by red, green and blue) and this partition divides the geometric domain into three segments with almost equal ($\pm1$) number of octants (see \cite{BursteddeWilcoxGhattas11}). }
	\vspace{10mm}
	\label{spacefillingcurve}
\end{figure}

1. \textbf{Space-filling curve.} Octants within an octree can be assigned a natural ordering by traversing across all leaves. This results in a one-dimensional sequence corresponding to a space-filling Z-shaped curve in the geometric domain. p4est extends this concept to a forest of octrees by connecting the space-filling curve between octrees, thus generating a total ordering of all octants in the domain (see Figure \ref{spacefillingcurve}).

2. \textbf{Parallel partition.} The objective of the partition algorithm is to equidistribute the computational work uniformly among processes, which is necessary to ensure parallel scalability of applications. In addition to uniformly distributing octants, p4est allows for a user-defined weight function that returns a non-negative integer weight for each octant and creates a partition that is evenly distributed by weight. After the partition, the relative order of octants with respect to processes stays the same as the order in the space-filling curve.

3. \textbf{Ghost.}
p4est is parallelized using the Message Passing Interface (MPI), which enables its applications to run on distributed memory supercomputers. For many computational applications, it is useful to know information on remote octants. Ghost algorithm collects one layer of neighboring octants on remote processes, providing the complete parallel neighborhood information. This feature reduces the communication pattern in the application to simple point-to-point transfers of numerical information.

The massively parallel LP code stores the particle data in a single array. Elements in the array contain all information about particles, such as coordinates, velocity, hydrodynamic states and supporting quantities. Not only does this make the code easier to maintain and extend, but also it simplifies the work to communicate particle data among processes as the particle data is stored on a contiguous memory block. The particle data is uniformly distributed among available processes. Particles on local processes are sorted by following the order of their owner octants in the space-filling curve. Particles in the same octants are not imposed with any specific orders. 

The LP code has been parallelized using MPI. The macro-structure of the octrees and some global information, such as boundary conditions and initial states, are shared among processes. The micro-structure of octrees, i.e., the division of each octree into octants, particle data and some other highly memory-demanding data structures designed for certain problems are dynamically managed and distributed among all processes. In order to achieve balanced computational loads across processes, we assign weights in the p4est {\it Partition} algorithm as the number of computational particles in octants. As a result, local numbers of octants  vary among processes if the particle distribution is non-uniform, but computational particles are evenly distributed among processes. This partition optimizes the load balance as the neighbor searching and stencil operations dominate the computational time. 

p4est allows computational domains to be composed of multiple connected octrees, referred to a forest of octrees.  Neighboring octrees can be arbitrarily rotated against each other, communicating through non-local connections. This flexible encoding of octrees enables  implementations of a wide range of possible computational domains and periodic boundary conditions without any special treatment to stencils.

The LP code strictly follows the idea of objected-oriented design principles, dividing the algorithm into major tasks and implementing them with classes. LP supports algorithm extensions using polymorphic classes and public interface. It consists of several main classes. {\bf Initializer} reads the input information from a formatted input file and saves data into class member variables. These variables are used as inputs to member functions of several other classes to generate initial particle distributions, particle states, EOS models, boundary conditions, octree structures, data analysis etc. 
{\bf Global\_data} stores the particle data, EOS model, boundary model, octree data structures and implements particle initialization routines, particle migration, particle neighbor searching and ghost particle generation algorithms. 
{\bf LP\_solver} implements the first order upwind scheme and the second order Lax-Wendroff scheme for 2D and 3D problems. {\bf Octree\_manager} implements wrapper functions for building, adapting, balancing, and partitioning octrees using p4est library functions. {\bf Particle\_viewer} outputs particle data stored on local processes in the VTK format. p4est-based implementation of another Lagrangian particle based code AP-Cloud (Adaptive Particle-in-Cloud) for optimal solutions of Vlasov-Poisson problems can be found in \cite{YuKumar2022}.

\subsection{\label{sec:verification}Verification and parallel scalability}

The Lagrangian Particle code has been extensively verified and validated using both classical hydrodynamic problems \cite{SamulyakWang2018} and applied problems of plasma physics relevant to thermonuclear fusion \cite{SamulyakYuan2021,HollmannNaitlho2022}.  Since the main novelty of  the algorithmic part of this paper is the parallelization of the Lagrangian particle code for distributed memory supercomputers, it is sufficient to show that the parallel code gives the same accuracy as the original code for some typical hydrodynamic problems. We use the classical Yee vortex problem to demonstrate the numerical accuracy and convergence. 

\begin{figure}[h!]
	\centering
	\subfigure[Pressure \label{yeep}]{\includegraphics[width=.8\linewidth]{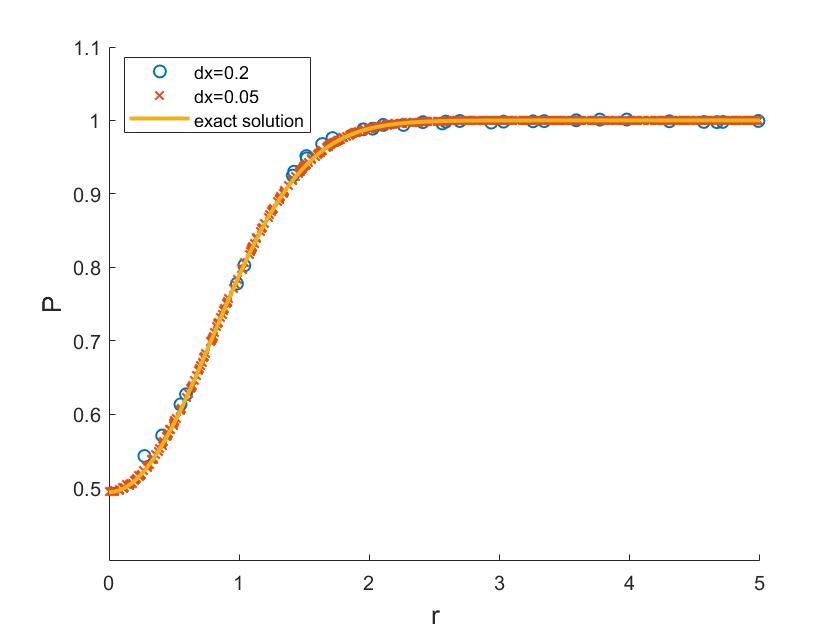}}
	\subfigure[Velocity \label{yeevelocity}]{\includegraphics[width=.8\linewidth]{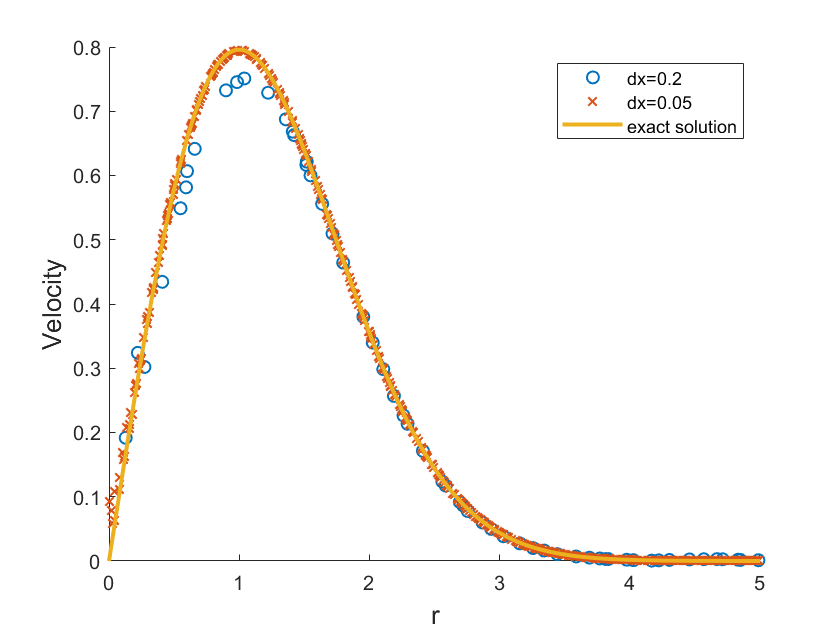}}
	\caption{Radial scatter plots of particles states of Yee vortex simulation at $t=8$}
	\label{yee}
\end{figure}

The Yee vortex \cite{YeeVinokur2000}  is an isentropic 2D vortex in a free-stream flow with the velocity $(u_{\infty}, v_{\infty})=(1,0)$, density $\rho_{\infty}=1$, and pressure $p_{\infty}=1$. The initial conditions for the Yee vortex are as follows
\begin{equation}
	\begin{split}
		&\rho=T^{\frac{1}{\gamma-1}}=[1-\frac{(\gamma-1)\beta^{2}}{8\gamma\pi^{2}}e^{1-r^{2}}]^{\frac{1}{\gamma-1}},  \\
		&   u = 1-\frac{\beta}{2\pi}e\frac{1-r^{2}}{2}(y-y_{0}), \\
		&  v=\frac{\beta}{2\pi}e^{\frac{1-r^{2}}{2}}(x-x_{0}),\\
		& P = \rho^{\gamma},
	\end{split}
\end{equation}
where $(x_{0},y_{0})$ are coordinates of the vortex center, $r^{2}=(x-x_{0})(x-x_{0})+(y-y_{0})(y-y_{0})$, and $\beta$ is the vortex strength. Notice that the vortex extends to the entire 2D space, but it can be truncated for all practical purposes as the velocity rapidly decays to the free-stream velocity. The exact solution with given initial states is just a passive convection of the vortex moving with the free-stream velocity. Therefore, it provides a good accuracy measure for numerical schemes for the nonlinear Euler equations. For the numerical test, we select $\beta=5$ and $\gamma=1.4$. Simulations that used a perfect hexagonal initial placement of particles and small random perturbations of initial particle coordinates produced very similar results in terms of numerical errors. The norms of numerical errors of pressure and velocity for the initial hexagonal placement of particles are given in Tables \ref{yeeperror} and \ref{yeeverror}.The convergence order is higher than second in both $L_{2}$ and $L_{\infty}$ norms. Figure \ref{yee} shows radial scatter plots for different numerical resolutions and visually demonstrates the numerical convergence. While the problem is two-dimensional, it has also been performed with the full-3D code by assuming constant states in the third dimension, thus testing the performance of data structure and functions relevant to octrees. 3D results were used for the parallel scalability analysis below.

\begin{table}[h!]
	\centering
	\begin{tabular}{ccccc}
		\toprule  
		dx&$L_{2}$&$L_{\infty}$&ratio&order \\ 
		\midrule  
		0.4&0.02508&0.1488&-&- \\
		\midrule  
		0.2&0.00549&0.0366&4.57&2.19\\
		\midrule  
		0.1&0.0011&0.0063&4.99&2.32\\
		\midrule
		0.05&0.00029&0.0017&3.77&1.91\\
		\bottomrule  
	\end{tabular}
	\caption{Errors of pressure at $t=8$ and convergence rates for 2D Yee vortex problem.}
	\label{yeeperror}
\end{table}

\begin{table}[h!]
	\centering
		\begin{tabular}{ccccc}
		\toprule  
		dx&$L_{2}$&$L_{\infty}$&ratio&order \\ 
		\midrule  
		0.4&0.067&0.369&-&- \\
		\midrule  
		0.2&0.01789&0.1035&3.75&1.91\\
		\midrule  
		0.1&0.00324&0.00241&5.52&2.47\\
		\midrule
		0.05&0.00056&0.00047&5.78&2.53\\
		\bottomrule  
	\end{tabular}
	\caption{Errors of velocity at $t=8$ and convergence rates for 2D Yee vortex problem.}
	\label{yeeverror}
\end{table}

\begin{figure}[h!]
	\centering
	\includegraphics[width=0.8\linewidth]{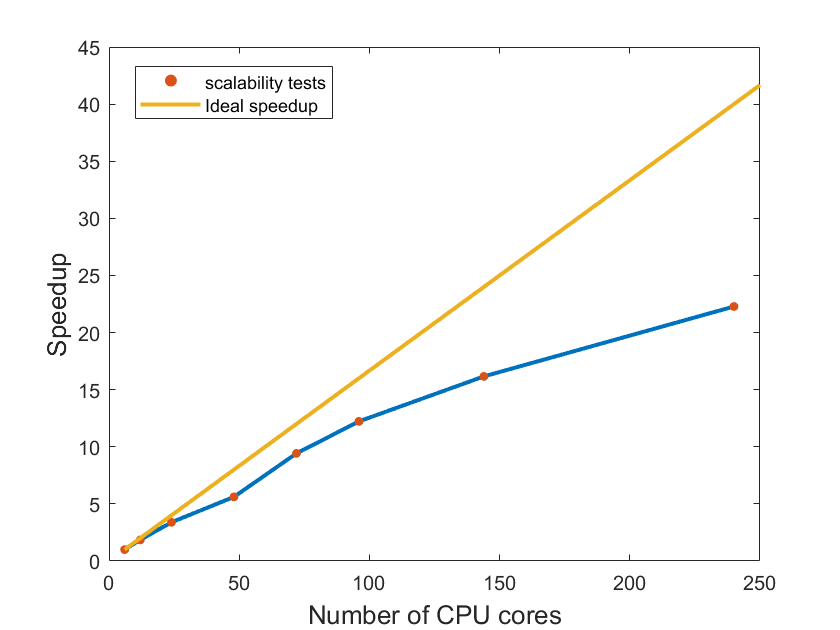}
	\caption{Strong scaling for massively-parallel LP code with using 1,060,000 particles. The largest speedup (22.28) was reaches at 240 CPU cores, which was 55.7\% of ideal speedup.}
	\label{strongscalingplot}
\end{figure}

We have studied parallel scalability of the code based on the Yee vortex problem in 3D using the Seawulf cluster located at Stony Brook University.  We tested the strong scaling by running the Yee vortex simulation with different number of CPU cores while keeping the total number of computational particles constant (1,060,000 particles were used in the strong scalability tests).  Figure \ref{strongscalingplot} shows the strong scalability of the LP code.  

\begin{figure}[h!]
	\centering
	\includegraphics[width=0.8\linewidth]{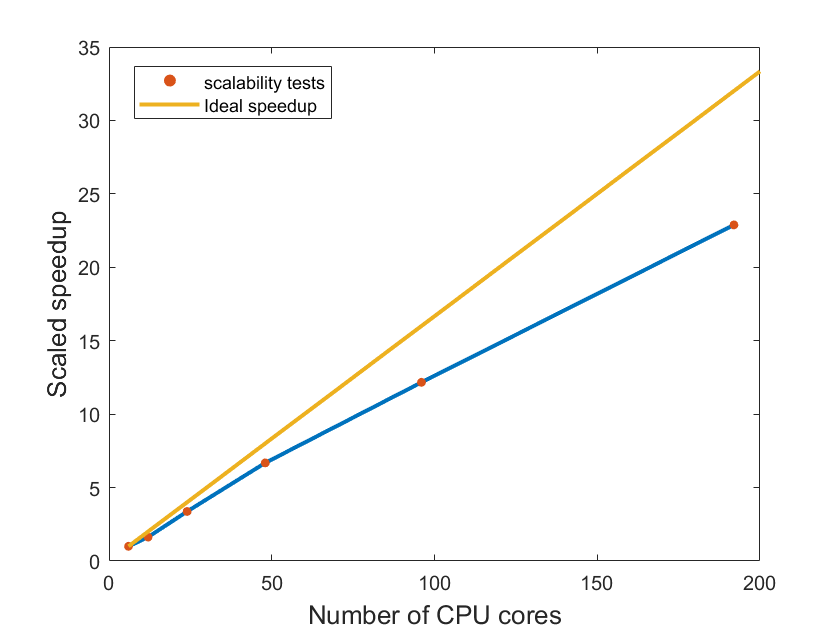}
	\caption{Weak scaling for massively-parallel LP code. The largest speedup (22.89) was reached at 192 CPU cores, which was 71.5\% of ideal speedup.}
	\label{weakscalingplot}
\end{figure}

The weak scaling was measured by running the code with different number of CPU cores and using a proportionally scaled number of particles such that the number of particle  per core remained constant. The results of weak scaling tests are demonstrated in Figure \ref{weakscalingplot}.

The tests demonstrated satisfactory scalability. As expected, the weak scaling outperformed the strong one. Since only the weak scaling is important for scientific computing in our area (which requires solving problems of ever-increasing size in a reasonable amount of time), achieving more than 70\% of ideal speedup and a linear scaling is sufficient for practical applications. It is important to note that the saturation of scaling has not been reached as we limited the test to 200 CPU cores, a typical allocation for our applications runs. We have reasons to expect that the weak scaling beyond 200 cores will remain close to linear.  

\section{Applications}

\subsection{\label{sec:pellets} Summary of thermonuclear fusion applications}

The Lagrangian particle code has been broadly applied to the simulation of impurities in hot thermonuclear fusion plasmas. Thermonuclear fusion devices called tokamaks explore the magnetic confinement of hot deuterium-tritium plasma for achieving conditions of thermonuclear fusion burn. The International Thermonuclear Experimental Reactor (ITER) [www.iter.org] is an international nuclear fusion research and engineering project that builds a new generation tokamak in Cadarache, France. Upon completion, the ITER tokamak is expected to achieve the fusion energy gain of at least 10 times larger compared to the energy deposited into the plasma. The ITER tokamak requires the injection of cryogenic material into the plasma for the purpose of fueling and plasma disruption mitigation. 

As a cryogenic pellet is injected into a thermonuclear plasma, it is hit by hot plasma electrons that rapidly ablate the pellet surface leading to the formation of a cold and dense ablation cloud. This ablated cloud absorbs most of the incoming plasma electrons. The fraction of electrons that reach the pellet surface is responsible for the pellet ablation rate. In our recent works,  2D axisymmetric  \cite{Bosviel2021} and full 3D simulations \cite{SamulyakYuan2021,HollmannNaitlho2022} of ablating pellets in tokamaks have been performed, and the influence of tokamak plasma conditions, magnetic field and 3D plasma drift effects on their ablation rates have been studied and compared with the available experimental data. The pellet ablation problem is intrinsically multiscale. The resolution of large gradients of plasma states occurring at sub-millimeter length scales near the pellet surface is critically important for computing pellet ablation rates. At the same time, large, device-scale expansion of the ablated material along tokamaks change the background plasma properties and affect the distribution of plasma heat sources for the pellet ablation. The adaptive property of the Lagrangian particle method made it possible to accurately resolve small scale processes near the pellet surface while propagating the ablated material over distances of dozens of meters. 

This section intended to emphasize adaptive multi-physics capabilities of the Lagrangian particle code. Detailes of the plasma physics simulations summarized above go beyond the scope of this paper and can be found in the referenced papers.  

\subsection{\label{sec:jets}Supersonic jets for LPWA}

In this section, we present applications of the Lagrangian Particle code to the problem of nozzle design optimization for laser plasma wakefield acceleration (LPWA) experiments being conducted at Brookhaven National Laboratory.   LPWA is potentially a high-impact research area. 
By replacing conventional, large-size accelerator structures with electromagnetic wakes created in plasma by lasers or high energy charged particles, LPWA is expected to increase the accelerating field gradients by orders of magnitude, leading to the reduction of accelerator sizes from the scale of tens of kilometers to the meter scale.

Broad studies of the plasma wakefield acceleration of electrons are being conducted at the Brookhaven National Laboratory (BNL) based on the powerful CO2 laser facility. Our research group supports this theoretical and experimental effort by numerical simulations. The main processes of interest, the CO2 laser interaction with hydrogen gas, ionization of hydrogen, formation of plasma wakes and the subsequent acceleration of electrons are described by electromagnetic system of equations for relativistic particles and plasma. The main simulation code is SPACE \cite{YuKumar2022}, a parallel, 3D electromagnetic Particle-in-Cell code with atomic physics support, developed in our research group. Recent simulations results are described in \cite{Kumar2019,Kumar2021}.

\begin{figure}[h!]
	\centering
	\includegraphics[width=.6\linewidth]{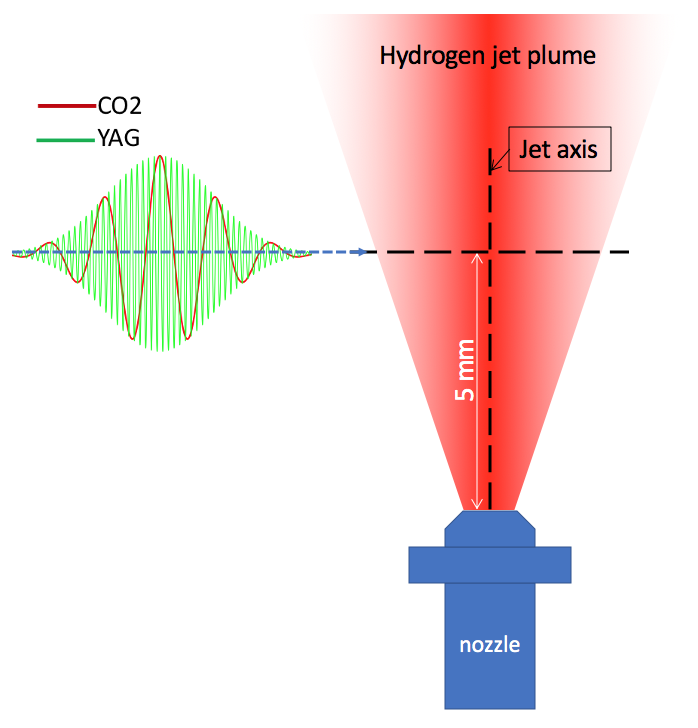}
	\caption{Schematic of experiment on laser-plasma wakefield acceleration. CO2 and YAG denote laser beams generated by the CO2 and solid state lasers, respectively, propagating across the hydrogen jet.}
	\label{fig:nozzlelaser}
\end{figure}

There is one component of the laser plasma wakefield experiments that is in the range of classical hydrodynamics. The problem is to obtain a proper distribution of hydrogen gas in vacuum before its interaction with the laser. Various designs have been considered for this tasks, including a hydrogen chamber with special windows for the laser beam. While various pro- and contra- of such chambers are being discussed, the BNL experiment currently operates with a simple solution - a hydrogen jet injected into the interaction region  through a nozzle connected to a  high-pressure reservoir. While being the simplest solution from the engineering point of view, the nozzle jet does not provide completely uniform initial gas distribution, which is desirable for the experiment. The purpose of numerical simulations is to optimize the process and obtain a nozzle  that provides the most uniform hydrogen density distribution of specific requirements. A schematic of the BNL laser plasma wakefield experiment is shown Figure \ref{fig:nozzlelaser}.

A nozzle is typically attached to a high-pressure reservoir containing the working gas. Computationally, we model the presence of the reservoir by special boundary conditions at the nozzle inlet. Three hypotheses are used in order to calculate the pressure, density and velocity of the particles entering the nozzle. 
\begin{itemize}
	\item The pressure of newly injected particles into the nozzle is equal to the average of the pressure of the particles in a small neighborhood of the nozzle inlet for the current timestep. (For the first timestep the pressure of the reservoir is used.)
	\item The entropy in the reservoir is equal to the entropy in the neighborhood of the nozzle inlet. In terms of pressure and density, this condition can be written as:
	\[ \frac{ P_{res}}{ \rho_{res}^{\gamma}} = \frac{ P_{inf}}{ \rho_{inf}^{\gamma}},
	\]
	where the subscripts 'res' and 'inf' denote the reservoir and nozzle inflow quantities.
	
	\item The transverse velocity component of particles is negligibly small (assumed zero), and the longitudinal velocity component is calculated using the enthalpy condition as
	\[ \frac{1}{2} u_{res}^2 + \frac{1}{\gamma -1} c^2_{res} = \frac{1}{2} u_{inf}^2 + \frac{1}{\gamma -1} c^2_{inf} \]
	where $c$ is the sound speed, given by $c^2 = \gamma \frac{P}{\rho}$. 
	Assuming $u_{res} = 0$, the velocity of the inflow particles can be calculated.  
\end{itemize}

\begin{figure}[h!]
	\centering
	\subfigure[Convergent-divergent nozzle \label{cd_nozzle}]{\includegraphics[width=0.8\linewidth]{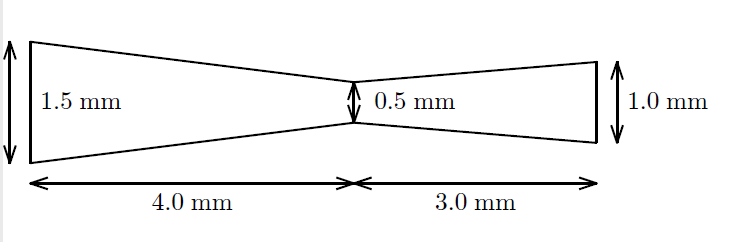}}
	\subfigure[Cylindrical-divergent nozzle \label{cyl_nozzle}]{\includegraphics[width=0.8\linewidth]{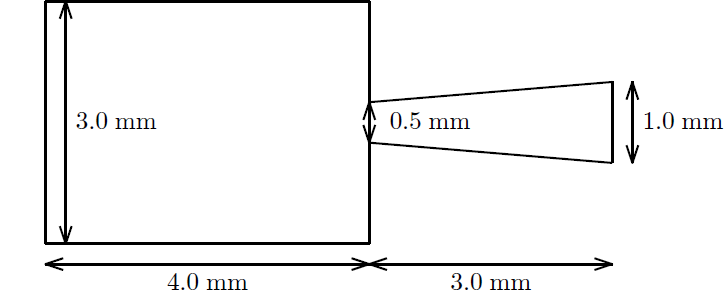}}
	\caption{Schematics and design parameters for convergent-divergent nozzle (a) and cylindrical-divergent nozzle (b)}
	\label{nozzle_params}
\end{figure}

\begin{figure}[h!]
	\centering
	\subfigure[Jet from convergent-divergent nozzle \label{jet_cd_nozzle}]{\includegraphics[width=1.\linewidth]{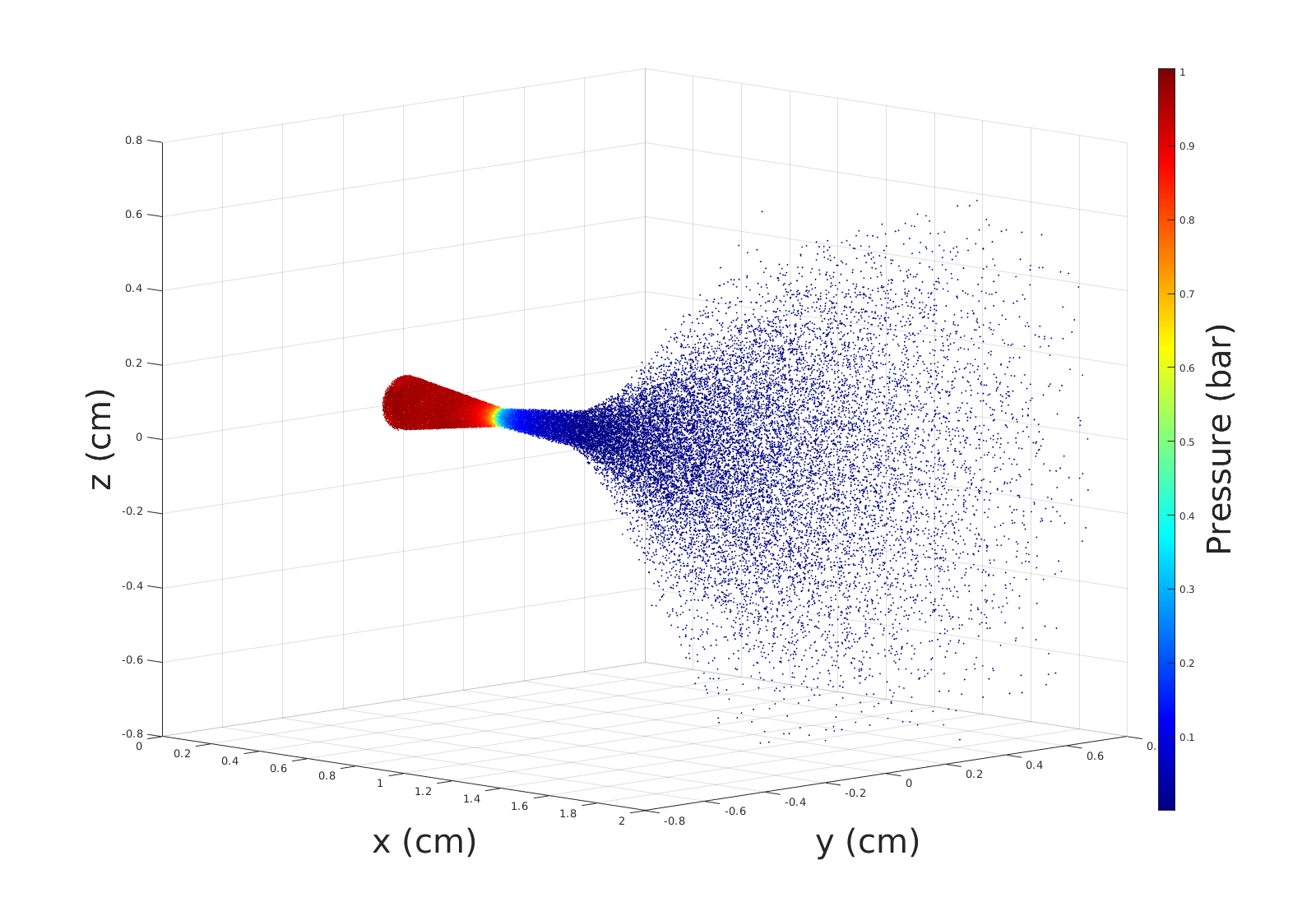}}
	\subfigure[Jet from cylindrical-divergent nozzle \label{jet_cyl_nozzle}]{\includegraphics[width=1.\linewidth]{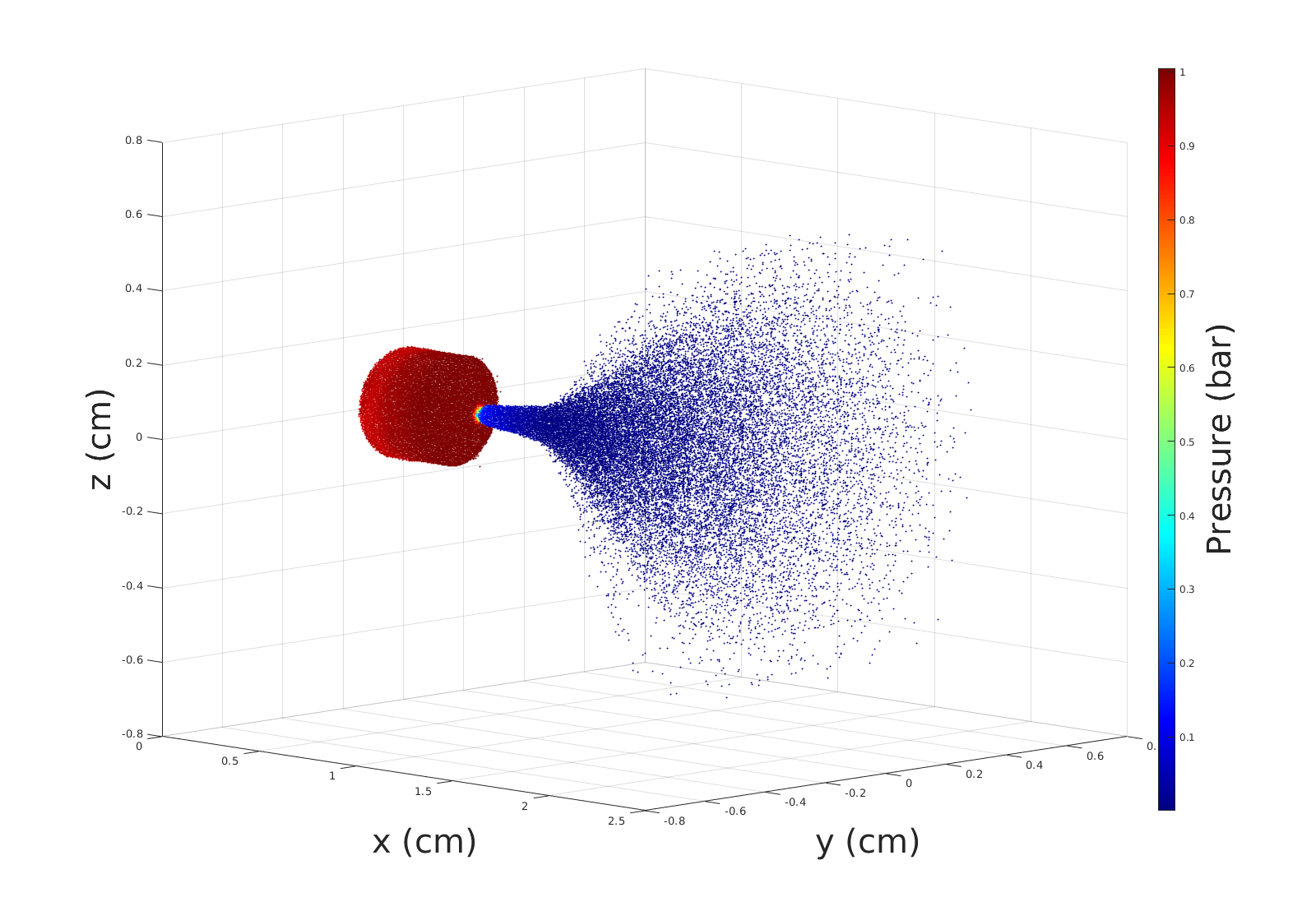}}
	\caption{LP simulation of hydrogen jets. Pressure distributions at $t = 50 \mu s$ are shown.}
	\label{ConDivNozzle}
\end{figure}

Two types of nozzles have been used in LPWA experiments, a cylindrical-divergent nozzle and a convergent-divergent one. Their schematics and geometric measurements are given in Figure \ref{nozzle_params}, and typical 3D numerical simulations are shown in Figures \ref{jet_cd_nozzle} and \ref{jet_cyl_nozzle}.

\begin{center}
	\begin{figure}[h!]
		\includegraphics[width=1.0\linewidth]{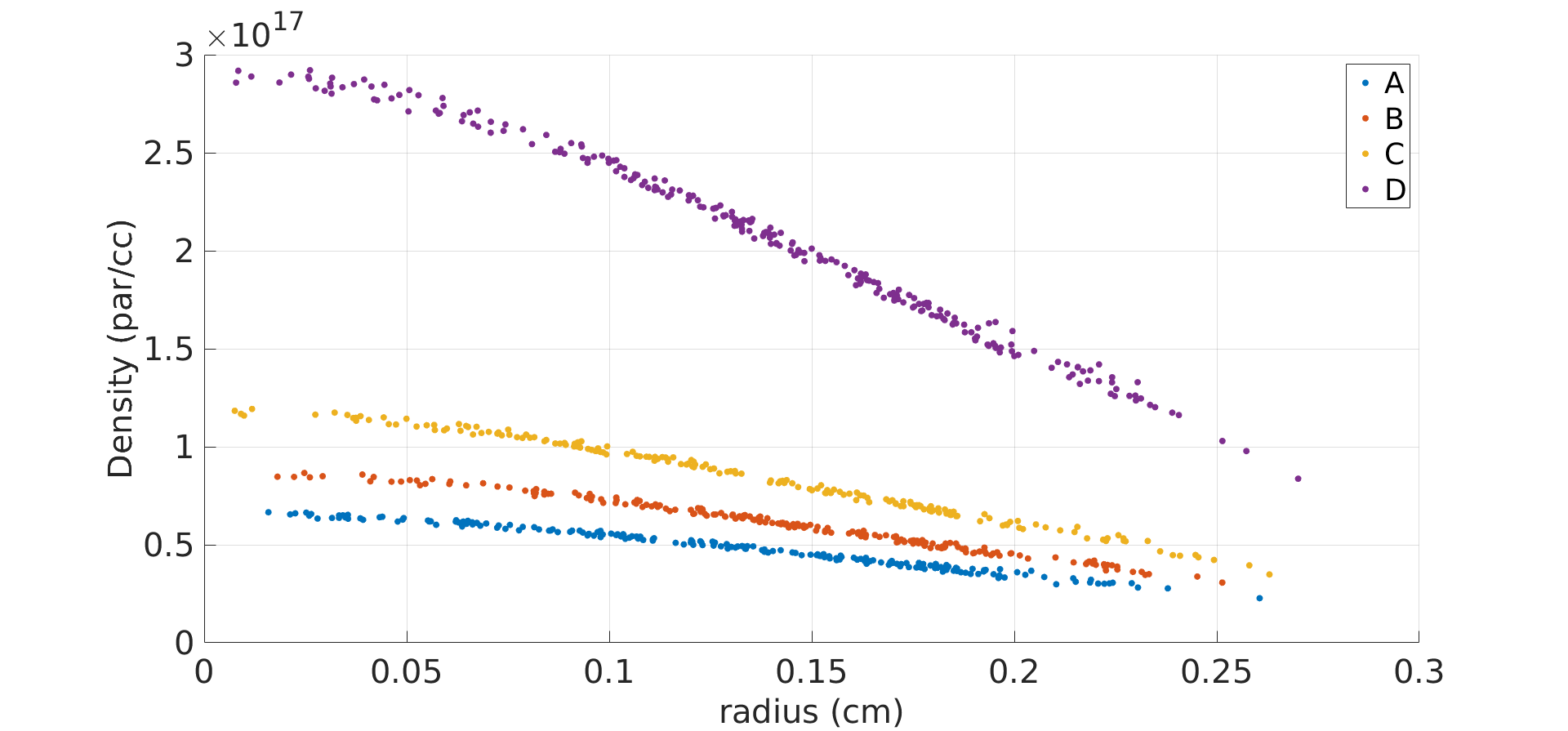}
		\caption{Radial profiles of hydrogen density at 5 mm form the nozzle exit for the reservoir temperature 293K and the following values of the reservoir pressure (bar): 1 (A), 1.27 (B), 1.72 (C), and 4.21 (D).}
		\label{CylDiv4casesDen}
	\end{figure}
\end{center}

The radial profiles of hydrogen density for the jets formed by the cylindrical-divergent nozzle at various reservoir parameters, used for determining plasma density after the interaction with laser, are shown in Figure \ref{CylDiv4casesDen}. Similar profiles were also obtained for the convergent-divergent nozzle.

\begin{figure}[h!]
	\centering
	\includegraphics[width=0.9\linewidth]{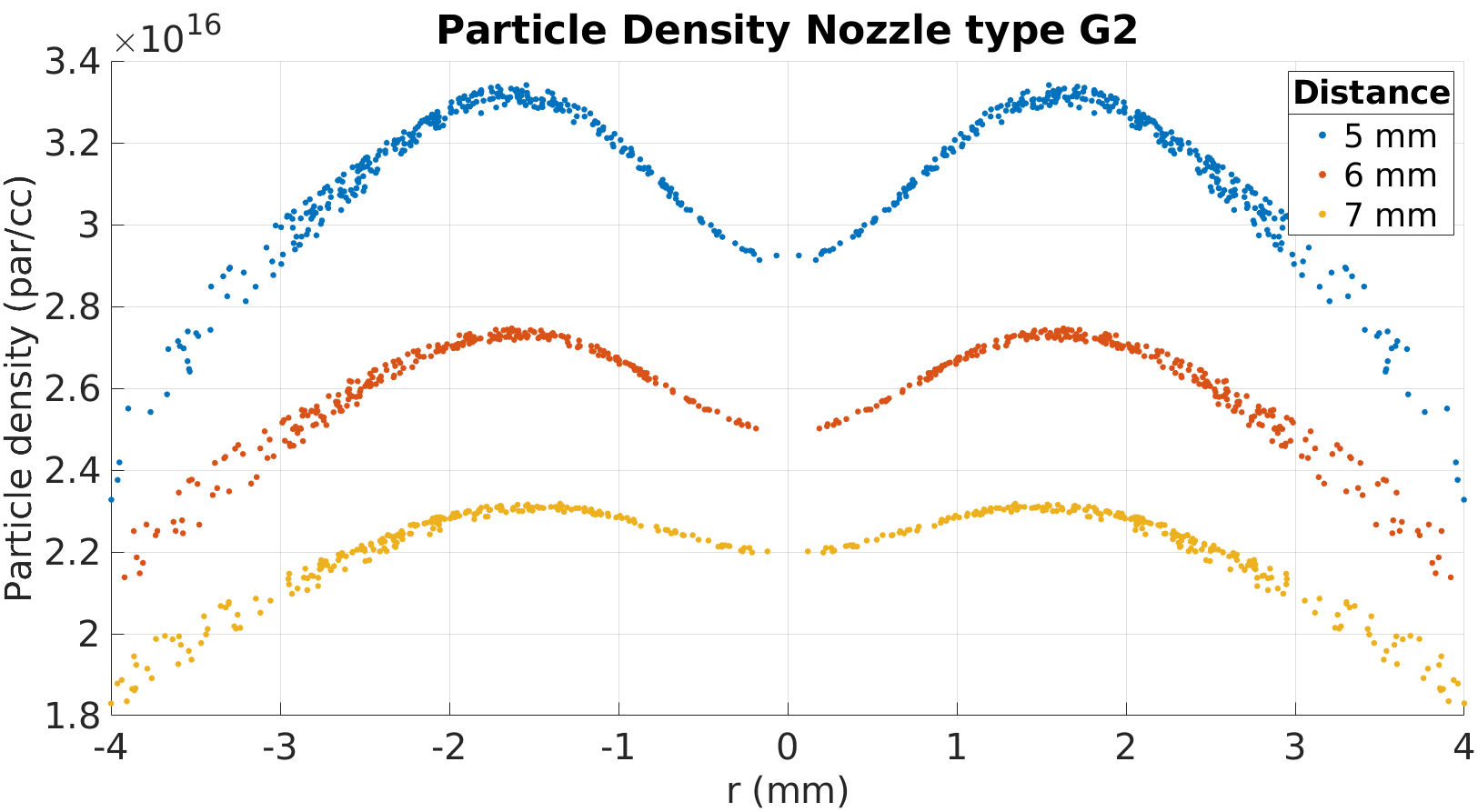}
	\caption{Hydrogen density for the optimized nozzle at several distances from the nozzle exit.}
	\label{fig:g2B}
\end{figure}

By optimizing the convergent-divergent nozzle shape, we were able to obtain reasonably small variations of hydrogen density over the distance of 6 mm as shown in Figure  \ref{fig:g2B}. The optimized nozzle dimensions are as follows: diameter of the reservoir opening is 1 mm, length of the convergent section is 4 mm, diameter of the neck is 0.5 mm, length of the divergent section is 8 mm, and diameter of the nozzle exit is 4.1 mm.

\section{Summary and conclusions}

In this paper, we described a parallel Lagrangian particle algorithm for compressible hydrodynamic problems and the corresponding object-oriented code for massively-parallel distributed memory supercomputers. The Lagrangian particle (LP) algorithm  implements an unsplit Lax – Wendroff type scheme with a limiter. In the location of points selected by a state estimator, the Lax–Wendroff scheme is blended with a first-order upwind method to control numerical dispersion. In the LP method, approximations of spatial derivatives are obtained by employing a local polynomial fit known also as the generalized finite difference (GFD) method. The LP method is second-order accurate in space and time and it is extendable to higher order accuracy.

The particle manager module of the massively-parallel LP code is based on the p4est library. p4est enables a dynamic management of a collection of adaptive octrees on distributed memory supercomputers. It has the functionality of building, refining, coarsening, 2:1 balancing and partitioning of computational domains composed of multiple connected two-dimensional quadtrees or three-dimensional octrees, referred to as a forest of octrees. The Lagrangian particle code has been verified, validated, and tested on distributed memory supercomputers. It demonstrated good strong and weak scalability on hundreds of supercomputer cores. 

The massively-parallel Lagrangian particle code is applicable to a variety of fundamental science and applied problems. Its current most important application is simulation of impurities in thermonuclear fusion devices. We briefly summarize the thermonuclear fusion applications in this paper and referred the reader to published works for further details.  Hydrodynamic simulations for the nozzle design optimization for supersonic hydrogen jets are described in more detail. Such simulations support experimental work at Brookhaven National Laboratory on CO2 laser-driven plasma wakefield acceleration. 
 
\vskip10mm

{\bf Acknowledgments}

This work was supported in part by the U.S. Department of Energy grants SciDAC Center for Tokamak Transient Simulations and Grant No. DE-SC0014043.

\bibliography{refs}

\begin{thebibliography}{10}
\expandafter\ifx\csname url\endcsname\relax
  \def\url#1{\texttt{#1}}\fi
\expandafter\ifx\csname urlprefix\endcsname\relax\def\urlprefix{URL }\fi
\expandafter\ifx\csname href\endcsname\relax
  \def\href#1#2{#2} \def\path#1{#1}\fi

\bibitem{Monaghan2005}
J.~Monaghan, Smoothed particle hydrodynamics, Rep. Prog. Phys. 68 (2005)
  1703--1759.

\bibitem{Frontiere2017}
N.~Frontiere, C.~Raskin, J.~Owen, {CRKSPH} conservative reproducing kernel
  smoothed particle hydrodynamics scheme, J. Comput. Phys. 332 (2017) 160--209.

\bibitem{Avesani2014}
D.~Avesani, M.~Dumbser, A.~Bellin, A new class of moving-least-squares
  {WENO-SPH} schemes, J. Comput. Phys. 270 (2014) 278--299.

\bibitem{Monaghan1992}
J.~Monaghan, Smoothed particle hydrodynamics, Annual Review of Astronomy and
  Astrophysics 30 (1992) 543--574.

\bibitem{Diltz1999}
G.~Diltz, Moving-least-squares particle hydrodynamics - {I}. consistency and
  stability, Int. J. Num. Methods in Engineering 44 (1999) 1115--1155.

\bibitem{Hopkins2014}
P.~Hopkins, {GIZMO}: a new class of accurate, mesh-free hydrodynamic simulation
  methods, Mon. Not, R. Astron. Soc. 450 (2014).

\bibitem{Liu1995}
W.~K. Liu, J.~Sukky, Y.~F. Zhang, Reproducing kernel particle methods,
  International Journal for Numerical Methods in Fluids 20 (1995) 1081--1106.

\bibitem{Oger2007}
G.~Oger, An improved {SPH} method: Towards higher order convergence, J. Comput.
  Phys. 225 (2007) 1472--1492.

\bibitem{SamulyakWang2018}
R.~Samulyak, X.~Wang, H.-C. Chen, Lagrangian particle method for compressible
  fluid dynamics, J. Comput. Phys. 362 (2018) 1--19.

\bibitem{BenitoUrena2001}
J.~Benito, F.~Urena, L.~Gavete, Influence of several factors in the generalized
  finite difference method, Appl. Math. Modelling 25~(12) (2001) 1039--1053.

\bibitem{SamulyakYuan2021}
R.~Samulyak, S.~Yuan, N.~Naitlho, P.~Parks, Lagrangian particle model for 3d
  simulation of pellets and {SPI} fragments in tokamaks, Nuclear Fusion 61~(4)
  (2021) 046007.

\bibitem{Zepeda2022}
M.~Zepeda, R.~Samulyak, Computational model for granular flow based on the
  lagrangian particle method, Computer Methods in Applied Mechanics and
  EngineeringSubmitted (2022).

\bibitem{Meagher1982}
D.~Meagher, Geometric modeling using octree encoding, Computer Graphics and
  Image Processing 19 (1982) 129--147.

\bibitem{BursteddeWilcoxGhattas11}
C.~Burstedde, L.~C. Wilcox, O.~Ghattas, {P4EST}: Scalable algorithms for
  parallel adaptive mesh refinement on forests of octrees, SIAM J. Sci. Comp.
  33~(3) (2011) 1103--1133.

\bibitem{YuKumar2022}
K.~Yu, P.~Kumar, S.~Yuan, A.~Cheng, R.~Samulyak, {SPACE}: 3d parallel solvers
  for vlasov-maxwell and vlasov-poisson equations for relativistic plasmas with
  atomic transformations, Comp. Physics Communications 277 (2022) 108396.

\bibitem{HollmannNaitlho2022}
E.~Hollmann, N.~Naitlho, S.~Yuan, R.~Samulyak, P.~Parks, D.~Shiraki,
  J.~Herfindal, C.~Marini, Measurement and simulation of small cryogenic neon
  pellet {Ne-I} 640 nm photon efficiency during ablation in {DIII-D} plasma,
  Phys. PlasmasSubmitted (2022).

\bibitem{YeeVinokur2000}
H.~C. Yee, M.~Vinokur, M.~J. Djomehri, Entropy splitting and numerical
  dissipation, Journal of Computational Physics 162~(1) (2000) 33--81.

\bibitem{Bosviel2021}
N.~Bosviel, P.~Parks, R.~Samulyak, Near-field models and simulations of pellet
  ablation in tokamaks, Physics of Plasmas 28~(1) (2021) 012506.
\newblock \href {https://doi.org/10.1063/5.0029721}
  {\path{doi:10.1063/5.0029721}}.

\bibitem{Kumar2019}
P.~Kumar, K.~Yu, R.~Zgadzaj, L.~Amorim, M.~Downer, J.~Welsh, V.~Litvinenko,
  N.~Vafaei-Najafabadi, R.~Samulyak, Simulation study of co2 laser-plasma
  interactions and self-modulated wakefield acceleration, Physics of Plasmas
  26~(8) (2019) 083106.

\bibitem{Kumar2021}
P.~Kumar, K.~Yu, R.~Zgadzaj, M.~Downer, I.~Petrushina, R.~Samulyak,
  V.~Litvinenko, N.~Vafaei-Najafabadi, Evolution of the self-injection process
  in long wavelength infrared laser driven lwfa, Physics of Plasmas 28~(1)
  (2021) 013102.

\end{thebibliography}

\end{document}